\documentclass[epj]{svjour}
\usepackage{epsfig,amssymb,amsmath,hyperref,bm}
\usepackage{graphics,graphicx,psfrag}

\def\beq{\begin{equation}}
\def\eeq{\end{equation}}
\def\bea{\begin{eqnarray}}
\def\eea{\end{eqnarray}}

\newcommand{\nonum}{\nonumber}

\newcommand\tJ{\tilde J}
\newcommand\tJp{\tilde J_{\perp}}

\begin{document}

\title{Ordering from frustration in a strongly correlated one-dimensional system}

\author{Siddhartha Lal
\inst{1}  
\and Mukul S. Laad
\inst{2}
}
\institute{The Abdus Salam ICTP. Strada Costiera 11, Trieste 34014, 
Italy \email{slal@ictp.it} \and 
Max-Planck-Institut f\"ur Physik Komplexer Systeme, 01187 Dresden, Germany 
\email{mukul@mpipks-dresden.mpg.de}}

\date{}


\abstract{We study a one-dimensional extended Hubbard model with longer-range 
Coulomb interactions at quarter-filling in the strong coupling limit.  
We find two different charge-ordered (CO) ground states as the strength 
of the longer range interactions is varied. At lower energies, these 
CO states drive two different spin-ordered ground states. A variety 
of response functions computed here bear a remarkable resemblance to 
recent experimental observations for organic TMTSF systems, and so 
we propose that these systems are proximate to a QCP associated with 
$T=0$ charge order. For a ladder system relevant to $Sr_{14}Cu_{24}O_{41}$, 
we find in-chain CO, rung-dimer, and orbital antiferromagnetic 
ordered phases with varying interchain couplings and superconductivity 
with hole-doping. RPA studies of many chains (ladders) coupled reveal 
a phase diagram with the ordered phase extended to finite temperatures 
and a phase boundary ending at a quantum critical point (QCP). Critical 
quantum fluctuations at the QCP are found to enhance the  
transverse dispersion, leading to a dimensional crossover and 
a $T=0$ decofinement transition.
}
\maketitle
%
%

\section{Introduction}
Electron crystallization, or the charge ordering of electrons due to 
interactions, is an issue of enduring interest in condensed matter physics
~\cite{mott,imada}.
A host of excellent studies clearly show the relevance of charge ordering (CO) in 
diverse systems like organics~\cite{org}, transition metal oxides~\cite{imada,salamon},
and coupled chain-ladder systems~\cite{mccaron}.  Generically, CO seems to compete 
with superconductivity~\cite{org,mccaron}, or with metallicity~\cite{imada,salamon}. 
Hence, in addition to its intrinsic academic interest, 
the study of the conditions favoring CO, along with its competition with 
metallic (magnetic) and/or superconducting states constitutes a problem of 
wide-ranging interest for a host of real systems.
\par
The simplest case of a CO state is the one where band fermions hop on a lattice
with a staggered (static) crystal potential~\cite{ashcroft}.  In this case, the 
CO gap is just the band-insulating gap.  Needless to say, this picture is too 
simple; it does not contain the ingredients necessary to study the coupled 
charge/spin aspects of the systems mentioned above: these point to the basic 
importance of the Hubbard $U$ in practice. In contrast to the band-CO case, one 
expects the correlation driven CO state to have drastic consequences for the 
magnetic order emerging at lower energy scales~\cite{imada,org,salamon,mccaron}, as well 
as for the competition between charge and spin orders manifested as one between 
Mott insulating (magnetic) and metallic (superconducting) states. 
\par
The availability of several powerful analytical as well as numerical methods 
in one dimension facilitates the detailed study of the effects of strong 
correlations in such systems. The study of the Hubbard model in one dimension, 
with and without 
extended Coulomb interactions, has a long history~\cite{giamarchi,capponi}. With 
the focus mostly on the case of $1/2$-filling~\cite{tsuchiizu}, 
systems away from $1/2$-filling have not received sufficient attention. 
Furthermore, studies at different fillings have typically concentrated on the 
weak coupling limit~\cite{yoshioka,tsuorig,yoshi1}, while real systems of 
interest~\cite{imada,org,salamon,mccaron} are generically in the strong coupling limit of 
appropriate Hubbard-type models. Additionally, one dimensional correlated 
models with couplings to the lattice have also been studied~\cite{riera,kuwabara,yoshi2}, 
revealing phase diagrams with various different kinds of charge orders (CO) and 
spin orders (SO). 
\par
The issue of whether various types of CO can, at fillings away from $1/2$, be driven 
purely by longer range Coulomb interactions remains, however, largely unadressed. 
The importance of extended Coulomb interactions was appreciated in an early work 
by Hubbard~\cite{hubbard}. Here, it was shown that the next
nearest neighbor Coulomb (nnn) interaction can be as large as 40 percent of the 
nearest neighbor (nn) interaction, with the intra-atomic Hubbard $U$ being the 
largest, and the one-particle hopping between neighbors the smallest energy 
scales in an effective one-band extended Hubbard model. In practice, coupling 
to high-energy Einstein (optical) phonons has the result of reducing the nn
coupling~\cite{kuwabara}; in a model with Coulomb interactions extended to next 
nearest neighbours, this gives rise to the possibility of tuning the ratio 
of nnn to nn interactions through $1/2$, a point at which interesting additional 
physics is expected. Thus, the search for competing charge ordered states driven 
purely by long-range electronic interactions, and the co-existence of charge and 
spin order, in a one dimensional $1/4$-filled electronic model are among the 
primary goals of the present work. Further, by working with a model in which 
the Coulomb correlations are much larger than the single particle hopping strength, 
we are able to obtain analytic results in a regime where little progress has been 
made. From our earlier discussion, our results are also clearly relevant to several 
material systems.    
\par
In the following section, we begin by presenting a derivation of a 1D transverse 
field Ising model (TFIM) effective pseudospin Hamiltonian in the extreme 
strong-coupling limit 
for the charge degrees of freedom of the $1/4$-filled 1D spinful electron Hubbard 
model with extended correlations. We then discuss the consequences various features 
of the phase diagram of the (exactly solvable) TFIM effective model have for 
the charge and spin degrees of freedom of the original electronic model. Various 
thermodynamic quantities of the electronic model, e.g., the optical conductivity, 
the dielectric function and the electronic contribution to the Raman scattering, 
are computed from the 
effective pseudospin theory. We follow this, in section III, with an analysis 
of the effective theories obtained for the case of a 2-leg ladder of 
two coupled TFIM chains with the inter-chain couplings in the strong coupling 
limit, as well as for a ladder doped with holes. In section IV, we then analyse 
the case of many such TFIM (chain and 
2-leg ladder) systems coupled to another by a random phase approximation (RPA) 
method~\cite{giamarchi,scalapino,schulz,carr}. The goal is to conduct a systematic study 
of the physics of dimensional crossover and deconfinement in such anisotropic 
strongly coupled lower dimensional systems, when coupled to one another. In 
section V, we present a comparison of our findings with some recent numerical 
works. Finally, we conclude in section VI. 

\section{Single chain in strong-coupling regime}
In this work, we study this issue within an extended quarter-filled 
Hubbard model on a linear chain, described by,
\bea
H_{\mathrm{eff}} &=& -t\sum_{i,\sigma}(C_{i\sigma}^{\dag}C_{i+1,\sigma}+h.c) 
+ (U-2zP)\hspace*{-0.1cm}\sum_{i}n_{i\uparrow}n_{i\downarrow}\nonumber\\
&&+ V\sum_{i}n_{i}n_{i+1}
+ P\sum_{i}n_{i}n_{i+2}
\eea
The couplings $U$, $V$ and $P$ represent the on-site Hubbard, nn
and nnn replusive interactions, $t$ represents the hopping 
strength and $z$ the coordination number for the lattice (which in the case of the 
linear chain is 2). This somewhat complicated looking microscopic model gives us 
the means to model extended Coulomb interactions in a $1/4$-filled system in the 
limit of strong coupling (as discussed below). Further, it is by now a well 
accepted paradigm that in theories of electrons 
in one-dimension interacting with one another through well-behaved (i.e., 
non-singular) potentials, the low-energy effective (fixed point) theory 
is one where the charge and spin degrees of freedom have separate dynamics
~\cite{giamarchi}. This is the concept of the Tomonaga-Luttinger liquid 
with spin-charge separation, and characterised by the spin fluctuations 
being those of an ideal $S=1/2$ XXX AF chain
\beq
H_{s} = J\sum_{j}{\bf S}_{j}.{\bf S}_{j+1},
\eeq
while the charge fluctuations 
are described by the Hamiltonian
\bea
H_{\mathrm{c}}&=&-t\sum_{i}(c_{i}^{\dag}c_{i+1}+h.c) +
(V-J/4)\sum_{i}n_{i}n_{i+1}\nonumber\\
&& + P\sum_{i}n_{i}n_{i+2}
\eea
that describes a model with frustrating interactions. The coupling $J$ is 
obtained from straightforward perturbation theory as $J\sim 4t^{2}/(U-V-4P)$.
In 1D, the projected fermions are spinless fermions with a hard-core constraint. 
\subsection{Effective TFIM model}
The model for spinless electrons given above has been considered as a model 
for studying the
effects of frustration on charge ordering~\cite{emery,kats}. 
For narrow-band systems, we consider the limit $t<<(V-J/4),P$.
Before embarking on a detailed derivation of the effective Hamiltonian in 
this regime, we present a heuristic one~\cite{emery}. For this, we employ 
an extension of the trick used for the $1d$ next-nearest neighbor Ising chain:
for $t=0$, we notice that with $(V-J/4)>2P$, the ground state is the usual CDW
(Wigner) crystal (and doubly degenerate, (..101010...) and (...010101...), 
for $1/4$-filling).  With $2P>(V-J/4)$, however, the dimerized state
(Peierls) is the ground state. The ground state has a degeneracy of four, with one 
of the states written schematically as (11001100....) and the other three achieved 
by shifting the state by one lattice site. Taking the state shown and
splitting this in a slightly different way, we have [...(01)(10)(01)(10)...].
Associating a pseudospin $\tau=1/2$ operator, with $\tau^{z}=+1$ for (10) and 
-1 for (01), the state is antiferromagnetic and doubly degenerate in terms of
the $\tau_{i}^{z}$. This state has a partner, obtained by shifting the state given 
above by 2 sites (or the pseudospin $\tau$ variables by one site). 
For small $t$, this is an attractive trick because (in
spin language) the transverse term does flip the $\tau_{i}^{z}$, but cannot 
break a pair.  This leads to the effective Hamiltonian~\cite{emery,kats},
\beq
H_{\mathrm{eff}}=-\sum_{l}[2t\tau_{l}^{x}+(V-J/4-2P)\tau_{l}^{z}\tau_{l+1}^{z}]
\eeq
\par
Let us now see how this simple Hamiltonian is obtained as the effective 
Hamiltonian for our original theory for the charge degrees of freedom in the 
limit of the couplings $U>>V,P>>t$.
We begin with the Hamiltonian for a spin chain system
\bea
H &=& -\sum_{n}[J_{x}(S_{n}^{x}S_{n+1}^{x} + S_{n}^{y}S_{n+1}^{y}) 
+ J_{1}S_{n}^{z}S_{n+1}^{z}\nonum\\ 
&& + J_{2}S_{n}^{z}S_{n+2}^{z} + hS_{n}^{z}]
\label{ham1}
\eea
where $S_{n}^{x}$, $S_{n}^{y}$ and $S_{n}^{z}$ and spin-1/2 operators. 
The couplings $J_{x}$, $J_{1}, J_{2}>0$ are the nearest neighbour (nn) XY, the 
nearest neighbour Ising and the next nearest neighbour (nnn) Ising couplings 
respectively and $h$ is the external magnetic field. For $h=0$, this 
Hamiltonian can be derived from the on-site Hubbard interaction 
$U\rightarrow\infty$ strong-coupling limit of an extended Hubbard 
model at 1/4-filling and with nn ($J_{1}\equiv(V-J/4)$) and nnn ($J_{2}\equiv2P$) 
density-density interaction couplings and the nn electron hopping ($J_{x}\equiv t$) 
via a Jordan-Wigner transformation (from a model of spinless-fermions to spins)
\beq
H = \sum_{n}[-\frac{J_{x}}{2}(c_{i}^{\dagger}c_{i+1}+h.c) 
+ J_{1}n_{i}n_{i+1} + J_{2}n_{i}n_{i+2}]~. 
\label{fermi}
\eeq
We study the problem in the 
limit of strong-coupling where $J_{1}, J_{2} >> J_{x}$ (but where 
$(J_{1}-2J_{2})\sim 2J_{x}$).
\par
Let us begin by studying the case of $J_{x}=0$~\cite{emery} (we 
will be studying eq.(\ref{ham1}) for the case of $h=0$ in all that 
follows). It is easy to see that for the case of $J_{1}>2J_{2}$, the 
ground state of the system is given by a Neel-ordered antiferromagnetic 
(AF) state with two degenerate ground-states given by
\bea
|AFGS1\rangle &=& |\ldots + - + - + - + - \ldots\rangle\nonum\\
|AFGS2\rangle &=& |\ldots - + - + - + - + \ldots\rangle
\label{wig1}
\eea
where we signify $S_{n}^{z}=1/2,-1/2$ by $+$ and $-$ respectively and we 
have explicitly shown the spin configuration in the site nos. 
$-3\leq n \leq 4$ in the ground states. In the original electronic 
Hamiltonian eq.(\ref{fermi}), this AF order corresponds to a Wigner 
charge-ordering (CO) 
in the ground-state. Similarly, for the case of $J_{1}<2J_{2}$, the 
ground state of the system is given by a dimer-ordered  
$(2,2)$ state \cite{emery} with four degenerate ground-states given by
\bea
|22GS1\rangle &=& |\ldots - + + - - + + - \ldots\rangle\nonum\\
|22GS2\rangle &=& |\ldots - - + + - - + + \ldots\rangle\nonum\\
|22GS3\rangle &=& |\ldots + - - + + - - + \ldots\rangle\nonum\\
|22GS4\rangle &=& |\ldots + + - - + + - - \ldots\rangle\nonum\\
\label{pei1}
\eea
where we signify $S_{n}^{z}=1/2,-1/2$ by $+$ and $-$ respectively and we 
have explicitly shown the spin configuration in the site nos. 
$-2\leq n \leq 5$ in the ground states. In the original electronic 
Hamiltonian eq.(\ref{fermi}), this $(2,2)$ order corresponds to a 
Peierls CO in the ground-state. Further, as the Hamiltonian 
eq.(\ref{ham1}) with $J_{x}=0=h$ is the one-dimensional axial next nearest 
neighbour Ising (ANNNI) model, we know that the ``frustration" point at 
$J_{1}=2J_{2}$~\cite{peschel} is gapless, highly degenerate and has 
a ground state entropy per site of $\ln (1+\sqrt{5})/2 \simeq 0.4812$ 
while the ordered ground states on either side of the frustration 
point have zero ground state entropy~\cite{liebmann}.
\par
We now work out the effect of the XY terms in the Hamiltonian (\ref{ham1}) 
on these ground states. Let us start with noting the effect of a XY term 
on a single nn spin-pair on the 4 degenerate $(2,2)$ ground-states; for 
purposes of brevity, we will denote the entire 
$J_{x}(S_{n}^{x}S_{n+1}^{x} + S{n}^{y}S_{n+1}^{y})$ 
term simply as $J_{x}^{n,n+1}$. Thus, 
\bea
J_{x}^{0,1}|22GS1\rangle &=& J_{x}^{0,1}|\ldots + - \ldots\rangle 
= \frac{J_{x}}{2}|\ldots - + \ldots\rangle\nonum\\
J_{x}^{0,1}|22GS2\rangle &=& J_{x}^{0,1}|\ldots + + \ldots\rangle 
= 0\nonum\\
J_{x}^{0,1}|22GS3\rangle &=& J_{x}^{0,1}|\ldots - + \ldots\rangle 
= \frac{J_{x}}{2}|\ldots + - \ldots\rangle\nonum\\
J_{x}^{0,1}|22GS1\rangle &=& J_{x}^{0,1}|\ldots - - \ldots\rangle 
= 0\nonum\\
J_{x}^{-1,0}|22GS1\rangle &=& J_{x}^{-1,0}|\ldots + + \ldots\rangle 
= 0\nonum\\
J_{x}^{-1,0}|22GS2\rangle &=& J_{x}^{-1,0}|\ldots - + \ldots\rangle 
= \frac{J_{x}}{2}|\ldots + - \ldots\rangle\nonum\\
J_{x}^{-1,0}|22GS3\rangle &=& J_{x}^{-1,0}|\ldots - - \ldots\rangle 
= 0\nonum\\
J_{x}^{-1,0}|22GS4\rangle &=& J_{x}^{-1,0}|\ldots + - \ldots\rangle 
= \frac{J_{x}}{2}|\ldots - + \ldots\rangle~.
\label{flip1}
\eea
\par
In a similar manner, we study the action of the operator $J_{x}^{n,n+1}$ 
on the two degenerate ground states of the AF ordered configuration as
\bea
J_{x}^{0,1}|AFGS1\rangle &=& J_{x}^{0,1}|\ldots + - \ldots\rangle 
= \frac{J_{x}}{2}|\ldots - + \ldots\rangle\nonum\\
J_{x}^{0,1}|AFGS2\rangle &=& J_{x}^{0,1}|\ldots - + \ldots\rangle 
= \frac{J_{x}}{2}|\ldots + - \ldots\rangle\nonum\\ 
J_{x}^{-1,0}|AFGS1\rangle &=& J_{x}^{-1,0}|\ldots - + \ldots\rangle 
= \frac{J_{x}}{2}|\ldots + - \ldots\rangle\nonum\\
J_{x}^{-1,0}|AFGS2\rangle &=& J_{x}^{-1,0}|\ldots + - \ldots\rangle 
\hspace*{-0.1cm}= \frac{J_{x}}{2}|\ldots - + \ldots\rangle~. 
\label{flip2}
\eea
\par
Defining bond-pseudospins $\tau_{i}^{z} = (S_{i}^{z} - S_{i-1}^{z})/2$, 
$\tau_{i}^{+} = S_{i}^{+}S_{i-1}^{-}$ and 
$\tau_{i}^{-} = S_{i}^{-}S_{i-1}^{+}$ (which can be rewritten in terms 
of fermionic operators in the original electronic Hamiltonian 
eq.(\ref{fermi}) as
$\tau_{i}^{z} = (n_{i} - n_{i-1})/2$, $\tau_{i}^{+} = c_{i}^{\dagger}c_{i-1}$ 
and $\tau_{i}^{-} = c_{i}c_{i-1}^{\dagger}$ respectively), we can 
write the four degenerate ground states of the $(2,2)$ ordered 
configuration in terms of these bond pseudospins as
\bea
|22GS1\rangle &=& |\ldots~ 0~  -~  0~  +~  0 \ldots\rangle\nonum\\
|22GS2\rangle &=& |\ldots\hspace*{0.1cm}  +~  0~  -~  0~  + 
\ldots\rangle\nonum\\
|22GS3\rangle &=& |\ldots~ 0~  +~  0~  -~  0 \ldots\rangle\nonum\\
|22GS4\rangle &=& |\ldots\hspace*{0.1cm}  -~  0~  +~  0~  - \ldots\rangle~,
\label{pei2}
\eea
where we have denoted $\tau_{n}^{z}=1/2$ as $+$ and $\tau_{n}^{z}=-1/2$ as 
$-$ and have explicitly shown the pseudospin configurations on the 
bond nos. $0\leq n \leq 4$. We can clearly see from eq.(\ref{pei2}) 
that these four 
ground-states break up into two pairs of doubly degenerate (AF) orderings 
of the pseudospins defined on the odd bonds ($|22GS1\rangle$ and 
$|22GS3\rangle$) on the even bonds ($|22GS2\rangle$ and 
$|22GS4\rangle$) respectively. It is also simple to see from eq.(\ref{flip1}) 
that the 
action of the operator $J_{x}^{n-1,n}$ (for the nearest neighbour pair 
of sites given by $(n-1,n)$) on these four ground states is to flip 
a pseudospin defined on the bond $n$ (lying in between the pair of 
sites $(n-1,n)$) or to have no effect at all. 
\par
We can now similarly see that the two-degenerate ground states of the AF 
ordered configuration can be written in terms of the bond-pseudospins 
defined above as
\bea
|AFGS1\rangle &=& |\ldots  +  -  +  -  + \ldots\rangle\nonum\\
|AFGS2\rangle &=& |\ldots  -  +  -  +  - \ldots\rangle
\label{wig2}
\eea
where we have explicitly shown the pseudospin configurations on the 
bond nos. $0\leq n \leq 4$. From eq.(\ref{wig2}), we see that the 
two degenerate 
ground-states have antiferromagnetic ordering of pseudospins on nn bonds; 
this can equally well be understood in terms of the ferromagnetic ordering 
of pseudospins on the odd bonds and on the even bonds separately. Further, 
from eq.(\ref{flip2}), we can see 
that the action of the operator $J_{x}^{n-1,n}$ (for the nearest 
neighbour pair of sites given by $(n-1,n)$) on these two ground states 
is again to flip a pseudospin defined on the odd (even) bond $n$ 
(lying in between the pair of sites $(n-1,n)$) against a background of 
ferromagnetically ordered configuration of pseudospins defined on the 
odd (even) bonds.
\par
Thus, we can model these pseudospin ordered ground states 
(\ref{pei2}),(\ref{wig2}) as well as {\it all} possible 
pseudospin-flip excitations above them (as given by action of  
operators of the type $J_{x}^{n-1,n}$ (\ref{flip1}),(\ref{flip2})) 
with the effective Hamiltonian
\bea
H &=& -\sum_{n\in odd}[2J_{x}\tau_{n}^{x} 
+ (J_{1}-2J_{2})\tau_{n}^{z}\tau_{n+2}^{z}]\nonum\\
&& -\sum_{n\in even}[2J_{x}\tau_{n}^{x} 
+ (J_{1}-2J_{2})\tau_{n}^{z}\tau_{n+2}^{z}]\nonum\\
&=& -\sum_{n}[2J_{x}\tau_{n}^{x} 
+ (J_{1}-2J_{2})\tau_{n}^{z}\tau_{n+2}^{z}]~,
\label{ham2}
\eea
where $n$ is the bond index. 
\par
  This is just the Ising model in a transverse field, which is exactly 
solvable~\cite{lieb,niemeijer} and has been studied extensively in 1D
~\cite{sachdev,chakrabarti}.  If $(V-J/4-2P)>0$, the ground state is 
ferromagnetically ordered in $\tau^{z}$, i.e, it corresponds to a Wigner CDW.
 For $(V-J/4-2P)<0$, the Peierls dimer order results in the ground state.  At 
$(V-J/4-2P)<2t$, the quantum disordered phase
has short-ranged pseudospin correlations, and is a charge
``valence-bond" liquid.  
The quantum critical point at $(V-J/4-2P)=2t$ separating these phases is a
deconfined phase with gapless pseudospin ($\tau$) excitations, and 
power-law fall-off in 
the pseudospin-pseudospin correlation functions.  Correspondingly, 
the density-density 
correlation function has a power-law singular behavior at low energy, 
with an exponent
$\alpha=1/4$ characteristic of the $2D$ Ising model at criticality.
The gap in the pseudospin spectrum
on either side of the critical point is given by $\Delta_{\tau} = 2|V-J/4-2P-2t|$.  
Further, the quantum-critical behaviour extends to temperatures as
high as $T\sim \Delta_{\tau}/2$~~\cite{kopp} and undergoes finite-temperature
crossovers to the two gapped phases at $T\sim |\Delta_{\tau}|$.
For $P=0$, the metallic phase for $V-J/4\le2t$ is a Luttinger liquid,
and in this limit, the low-energy physics is qualitatively similar to that of 
the usual $t-J$ model.  The ``Mott" insulating state for $V-J/4>2t$
 has Wigner CO in the 
ground state, and the M-I transition is of the Kosterlitz-Thouless
 type~\cite{tsvelik}.
\begin{figure}[htb]
\begin{center}
\scalebox{0.32}{
\psfrag{1}[bl][bl][4][0]{$T$}
\psfrag{2}[bl][bl][4]{$g$}
\psfrag{3}[bl][bl][3.5][0]{$g=0$}
\psfrag{4}[bl][bl][3.5][0]{Low $T$ gapped}
\psfrag{5}[bl][bl][3.5][0]{CO state} 
\psfrag{6}[bl][bl][3.5][0]{with LRO}
\psfrag{7}[bl][bl][4][0]{Quantum} 
\psfrag{8}[bl][bl][4][0]{Critical}
\psfrag{9}[bl][bl][3.5][0]{Low $T$ gapped}
\psfrag{10}[bl][bl][3.5][0]{short ranged} 
\psfrag{11}[bl][bl][3.5][0]{charge VB liquid}
\includegraphics{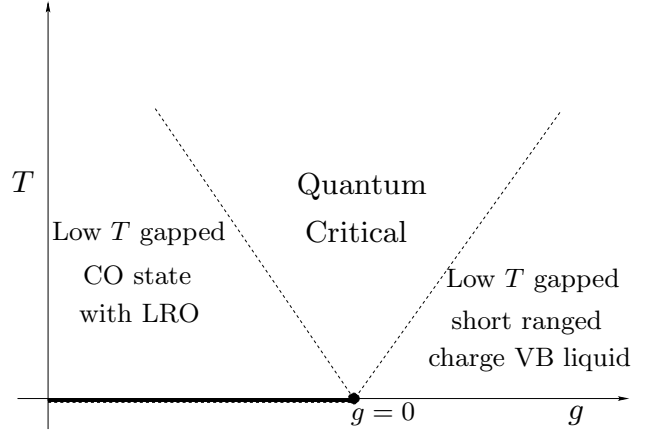}
}
\end{center}
\caption{A schematic phase diagram of the temperature $T$ vs. coupling 
$g=|(V-J/4-2P-2t)/(V-J/4-2P)|=\Delta_{\tau}/(V-J/4-2P)$ of the effective 1D Transverse Field Ising Model 
theory for the charge sector of our original electronic model. The black circle 
with $g=0$ represents the $T=0$ 
quantum critical point (QCP) separating an phase with true LRO given by Wigner 
CDW or Peierls dimer order 
(depending on whether $V-J/4>2P$ or $V-J/4<2P$ respectively, thick dark line) and 
a quantum disordered phase corresponding to a charge valence bond (VB) liquid. 
The finite-$T$ physics of the quantum critical region lying just above the QCP 
is discussed in subsection IIB. The dashed regions represent finite-$T$ crossovers 
to low-$T$ gapped charge ordered and charge VB liquid phases with no long-range 
order (LRO).}
\label{tfimphasediag}
\end{figure}
\par 
  The full Hamiltonian in our case for the strong-coupling limit is now 
given by
\beq
H_{\mathrm{eff}}=\hspace*{-0.1cm}-\hspace*{-0.1cm}\sum_{l}[2t\tau_{l}^{x}+(V\hspace*{-0.1cm}-\hspace*{-0.1cm}J/4\hspace*{-0.1cm}-\hspace*{-0.1cm}2P)\tau_{l}^{z}\tau_{l+1}^{z}]
+ J\sum_{l}{\bf S_{l}}.{\bf S_{l+1}}  
\eeq

  To study the magnetic phases, we adapt the Ogata-Shiba~\cite{ogata}
technique for our case.  This is possible if $J(\sim t^{2}/(U-V-4P))<<t,V$, 
in which case, the 
pseudospin part is first solved exactly (this is possible because of the 
known exact solution of the 1D transverse field Ising model), and the exchange
part is then treated as a perturbation.  Writing the total wavefunction as a 
product of a spin and pseudospin wavefunction (where the spin wavefunction is
defined, for a system of size $N$, in a Hilbert space of dimension $2^{N}$), 
i.e, $|\psi>=|\tau>\otimes|{S}>$,
and following standard degenerate perturbation theory, the spin degeneracy is 
lifted by the correction (of order $1/N$):
\beq
\hspace*{-0.5cm} <H_{\mathrm{eff}}>' = -2t<\tau^{x}>
+ \sum_{l} \tilde{J}_{l,l+1}({\bf S_{l}}.{\bf S_{l+1}}-1/4)
\label{hcomplete}
\eeq
 where the average $<..>'$ denotes that the average is taken over the exact
ground state $|\tau>$ of the pseudospin part above, i.e. $<A>'=<\tau|A|\tau>$ 
and $\tilde{J}_{l,l+1}=\frac{1}{N}(V-\frac{J}{4}-2P)<\tau_{l}^{z}\tau_{l+1}^{z}>$. 
\par
An interesting fact now emerges: Wigner CO (FM order of $\tau$) results in 
an HAFM $S=1/2$ spin model with the Hamiltonian
$H_{\mathrm{s}}=\tilde{J}\sum_{i}{\bf S}_{i}.{\bf S}_{i+1}$. This gives rise to a 
gapless AF ground
state for the spin degrees of freedom.  The charge (holon) 
excitations have a gap $\Delta_{\tau}=2(V-J/4-2P-2t)$; this corresponds to 
a linear confining potential for
holons.  On the other hand, Peierls dimerization in the
charge sector (AF Neel order of $\tau$) gives rise to dimerization in the 
spin sector, with the Hamiltonian
$H_{\mathrm{s}}=\tilde{J}\sum_{i}[1+(-1)^{i}\delta]{\bf S}_{i}.{\bf S}_{i+1}$.
The bare value of the dimerisation is given found from the perturbation 
theory as proportional to the Ising pseudospin (charge dimer) gap, 
i.e., $\delta\sim\Delta_{\tau}/\tilde{J}$. However, as we shall discuss now, 
a continuum field theoretic description of this Hamiltonian shows 
that the dimerisation parameter $\delta$ is renormalised by quantum 
fluctuations, and scales as $\delta\sim(\Delta_{\tau}/\tilde{J})^{2/3}$. 
\par
  Translated into fermion variables, this yields a sine-Gordon problem with 
$\beta^{2}=2\pi$, and describes an instability to a {\it singlet} pinned
ground state commensurate with the Peierls CO setting in at higher energies.    
The elementary excitations are solitons carrying $S^{z}=\pm 1$.
Scaling theory predicts a dimer gap, 
$\Delta_{\tau} \simeq \delta^{2/3}\tilde{J}$.  
Exactly at $\beta^{2}=2\pi$, the SG model has just {\it two} $S^{z}=0$
breather excitations with opposite parity~\cite{haldane}, the lowest, even 
parity breather being degenerate with the $S^{z}=\pm 1$ soliton doublet,
forming a $S=1$ triplet, while the second odd-parity breather is a singlet 
with a gap, $\sqrt{3}\Delta_{\tau}$.
  It is important to notice that {\it both} charge and spin order
arise from long range Coulomb interactions, and do not involve an electron
phonon coupling mechanism.  
\subsection{Consequences of CO on thermodynamic quantities in the 
quantum critical regime}  
Let us consider the implications of having the CO state in the high-$T$ 
regime, where one could imagine the system to be effectively one-dimensional.
In particular, we want to look at the $\omega,T$ dependence of the various 
response functions at high-$T$.  Using the exact 
solution of the pseudospin model in 1D, the high $T$ (in the ``quantum critical"
regime) behavior can be explicitly derived~\cite{sachdev}.
In fact, near Ising criticality, the response function, $\chi(r)\simeq r^{-1/4}$
 where $r=(x^{2}+\tau^{2})^{1/2}$ (with the velocity $v$ set to unity).  This
relation is still valid away from criticality in the ``short range" region,
$r<<\Delta_{\tau}$, where $\Delta_{\tau}=2|V-J/4-2P-2t|$ is the pseudospini gap
(which corresponds, in our case, to the charge gap) of the 1D-TFIM. 
As we will see, in what follows,
knowing the microscopic form of the Ising pseudospin gap $\Delta_{\tau}$ is 
enough information for us to be able to compute a host of thermodynamic response 
functions of the sytem and be able to relate them to the couplings in our 
original microscopic theory of interacting electrons in one dimension. 
Using this asymptotic form, we have
\beq
\chi_{\mathrm{crit}}(0,\omega)=-\frac{\sin(2\pi\Delta)}{(2\pi T)^{2-4\Delta}}B^{2}(\Delta-iS,1-2\Delta)
\label{chiasym}
\eeq
where $S=\frac{\omega}{4\pi T}$, and $\Delta=1/16$ is the conformal dimension.
$B(x,y)$ is the beta function.
\par
In the quantum critical region, an illuminating form is
\beq
\chi(k,\omega)=\frac{\chi(0,0)}{1-i\omega/\Gamma_{\mathrm{R}}+k^{2}\xi^{2}-(\omega/\omega_{1})^{2}}
\label{chicrit}
\eeq
where the quantities $\Gamma_{\mathrm{R}}=(2\tan(\pi/16)k_{\mathrm{B}}T/\hbar)e^{-\Delta_{\tau}/k_{\mathrm{B}}T}$, $\omega_{1}=0.795(k_{\mathrm{B}}T/\hbar)$ and 
$\xi=\hspace*{-0.05cm}1.28(c\hbar/k_{\mathrm{B}}T)e^{\Delta_{\tau}/k_{\mathrm{B}}T}$,
 are determined solely by $T$ and the fundamental natural constants, as expected in the QC regime.  We stress once again that here, $\Delta_{\tau}=2(V-J/4-2P-2t)$ 
is the energy gap to charge excitations in the Wigner/Peierls CO states 
(Ising pseudospin gap) described above.
 This represents the collective charge susceptibility, and the optical conductivity follows directly from 
 $\sigma(\omega)=-i\omega\chi(0,\omega)$, giving,
\beq
\sigma(\omega)=\frac{\chi(0,0)}{\Gamma_{R}}\frac{\omega^{2}}{(1-\omega^{2}/\omega_{1}^{2})^{2}+(\omega/\Gamma_{\mathrm{R}})^{2}}
\label{opcond}
\eeq
\par
The frequency dependent dielectric function, $\epsilon(\omega)$, is obtained from
$\epsilon(\omega)=1+(4\pi i\sigma(\omega)/\omega)$, and the electronic 
contribution to the Raman scattering is estimated therefrom to be given by
$I_{\mathrm{R}}(\omega)=\mathrm{Im}(1/\epsilon(0,\omega))$, for light polarized along the chain 
axis.  
In terms of the charge susceptibility, this is simply,
\beq
I_{\mathrm{R}}(\omega)=\mathrm{Im}\frac{1}{\epsilon(\omega)}=\frac{\frac{4\pi\chi(0,0)}{\Gamma_{\mathrm{R}}}F(\omega,T)}{1+(\frac{4\pi\chi(0,0)}{\Gamma_{\mathrm{R}}})^{2}F^{2}(\omega,T)}
\label{ramanshape}
\eeq
where 
$F(\omega,T)=\frac{\omega}{(1-\omega^{2}/\omega_{1}^{2})^{2}+(\omega/\Gamma_{\mathrm{R}})^{2}}$.
\par
$\chi"(k,\omega)$ has its maximum value at $\omega_{m}=\omega_{1}-i(\omega_{1}^{2}/\Gamma_{\mathrm{R}})$, implying that the collective mode broadens and 
shifts to higher energy {\it linearly} in $T$ with increasing $T$ at high 
temperatures.  
 Further, the $T$-dependent damping rate of the 
collective mode correlates well with the relaxational peak seen in transport,
underlying their common origin.  In fact, the $dc$ resistivity is linear in $T$
at high $T$, with ``insulating" features showing up at lower $T$.  In our picture, these are collective
(longitudinal) bosonic
 charge-density modes in the high-$T$ quantum critical region
above an incipient CO transition (expected to occur at low $T$).  In fig.(1), 
we show the electronic Raman lineshape as a function of $\omega/T$.  The sharp
low energy peak corresponds to the collective charge density fluctuation mode
of the CO ground state.   
\begin{figure}[htb]
\begin{center}
\scalebox{0.8}{
\includegraphics{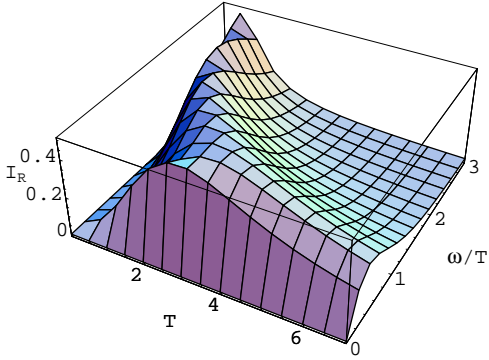}
}
\end{center}
\caption{A three-dimensional plot of the Raman intensity $I_{R}$ versus a 
scaled energy $\omega/T$ and temperature $T$ for parameter values of the 
original model which constitute a gap $\Delta_{\tau}=0.05 k_{\mathrm{B}}$.}
\label{raman}
\end{figure}
\par
  The corresponding frequency-dependent dielectric constant also shows an 
explicit $\omega/T$ scaling in the QC regime, or generally, at high-$T$,  
it shows strong $T$-dependence.  From fig.(2), we see 
that it becomes $\omega$-independent at high $T$, but appreciably increases as 
$T$ is lowered, with a maximum at $\omega\simeq T$. 
\begin{figure}[htb]
\begin{center}
\scalebox{0.8}{
\includegraphics{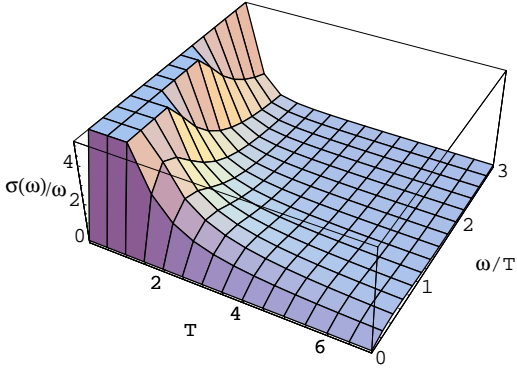}
}
\end{center}
\caption{A three-dimensional plot of $\sigma(\omega)/\omega$ versus a 
scaled energy $\omega/T$ and temperature $T$ for 
$\Delta_{\tau}=0.05 k_{\mathrm{B}}$. The limiting behaviors are:
(1) $\sigma(\omega) \simeq \frac{\omega^{2}}{T^{11/4}}$ for high $T$,
and $\sigma(\omega) \simeq \frac{1}{\omega^{3/4}}$ for low $T$.  
(2) $\epsilon'(\omega,T)=1+\frac{const}{T^{7/4}}$ for high $T$, 
and $\epsilon'(\omega,T) \simeq \frac{-1}{\omega^{7/4}}$ for low $T$. 
}
\label{sigmaom}
\end{figure}
\par
  The fact that organic charge transfer salts~\cite{bra} exhibit features very similar to those found 
above has interesting implications.  In light of our results, these anomalous features can now be identified with proximity to an underlying 
quantum critical point associated with charge (Wigner/Peierls) ordering. 
We recall that very recent work~\cite{mon,bishop} shows that the dimerized insulating state in TMTSF systems has charge order at low $T$.
Interestingly, $\epsilon'(0,\omega)$ indeed shows appreciable increase as $T$ is lowered, further supporting an interpretation based on
proximity to an underlying CO ground state~\cite{bra}.
 Hence, we conclude that observation of these features in
TMTSF systems constitutes strong evidence that this system is close to a putative QCP associated with charge
order.  Observation of dimerized/Neel ordered AFM states co-existing CO states at low $T$ are also naturally understood in light of 
the analysis above~\cite{bishop}.      

\section{Two-chain TFIM Ladder systems at strong inter-chain coupling}
We now consider the strong coupling version of a coupled two-chain ladder system, with
each chain being described by $H$ as in eq.(1).  In the strong coupling limit, where each
chain is described by a TFIM for charge degrees of freedom, the coupled chain model is 
constructed as follows.  
For $U\rightarrow\infty$, and $V,P>t$ (but $(V-J/4-2P)$ comparable to $t$),
the charge degrees of freedom
of the fermionic problem for each chain are described by an effective
pseudospin model on n-n bonds, via the effective Hamiltonian,
\beq
H^{chain} = -\sum_{j} [2t\tau_{j}^{x}
+ (V - J/4 - 2P)~\tau_{j}^{z}\tau_{j+1}^{z}]
\label{onechain}
\eeq
Rotating the pseudospin axis such that $\tau^{x}\rightarrow\tau^{z}$,
$\tau^{z}\rightarrow - \tau^{x}$ and coupling two such chains via
an interaction coupling $U_{\perp}$
and a two-electron interchain transfer $t_{\perp}$,
we have the effective Hamiltonian for the charge sector of the
two chain system as
\bea
H &=& -\sum_{j,a} [2t\tau_{j,a}^{z}
                + (V-J/4-2P)~\tau_{j,a}^{x}\tau_{j+1,a}^{x}]\nonumber\\
&&-\sum_{j,a,b\neq a} [U_{\perp}\tau_{j,a}^{z}\tau_{j,b}^{z}
+ ~t_{\perp}~ (\tau_{j,a}^{x}\tau_{j,b}^{x}
+ \tau_{j,a}^{y}\tau_{j,b}^{y})]~,
\label{twochain}
\eea
where $a,b=1,2$ is the chain index. Denote the in-chain pseudospin
coupling as $J = (V-J/4-2P)$ and the inter-chain pseudospin coupling
as $J_{\perp} = U_{\perp}$.
Here, we study the strong coupling version of this problem in several limits by 
deriving the respective low-energy effective Hamiltonians (LEEH).
The weak-coupling problem is studied elsewhere~\cite{next}.

\subsection{The case of $|J_{\perp}| >> |J|,~ t_{\perp}$}
For the case of $|J_{\perp}| >> |J|,~ t_{\perp}$~, the 2 chain system can be better
thought of as strongly-coupled rungs which are weakly coupled to their
neighboring rungs. Thus, we treat $J$ as a 
weak perturbation on the zeroth-order system of rungs defined by the large
coupling $J_{\perp}$, giving $H_{eff} = H_{0} + H_{1}$ where
\bea
H_{0} &=& -h\sum_{j,a} \tau_{j,a}^{z}
+ J_{\perp}\sum_{j,a,b\neq a}\tau_{j,a}^{z}\tau_{j,b}^{z}\nonumber\\*
H_{1} &=& - J\sum_{j,a} \tau_{j,a}^{x}\tau_{j+1,a}^{x}
- \frac{t_{\perp}}{2}\hspace*{-0.2cm}\sum_{j,a,b\neq a}\hspace*{-0.2cm}(\tau_{j,a}^{+}\tau_{j,b}^{-}
+ {\rm h.c})
\label{start}
\eea
where the effective magnetic field is given by $h=2t>0$.
\par
\subsubsection{LEEH for $J_{\perp}<0$}
For $J_{\perp} <0$ and
$h<< J_{\perp}$, we find that the triplet state
$|+\rangle = \frac{1}{\sqrt{2}}(|\uparrow\downarrow\rangle +
|\downarrow\uparrow\rangle)$ and the singlet state $|-\rangle = 
\frac{1}{\sqrt{2}}(|\uparrow\downarrow\rangle - 
|\downarrow\uparrow\rangle)$ are degenerate on any rung and 
are separated from 
all other states by a large gap of order $J_{\perp}$. Thus, these two states 
define the subspace which will determine the low-energy physics
of the system.
Identifying a pseudospin-1/2 operator $\xi_{j}$ with the low-energy
subspace on each rung, we treat the Hamiltonian $H_{1}$ as a
perturbation (to second order in $J/J_{\perp}$) to obtain the LEEH as
\bea
H &=& \sum_{j}[-\frac{J^{2}}{2J_{\perp}}
(\frac{J_{\perp}^{2} - 2h^{2}}{J_{\perp}^{2} - 4h^{2}})
\xi_{j}^{z}\xi_{j+1}^{z} - \frac{t_{\perp}}{2}\xi_{j}^{z}\nonumber\\*
&&\hspace*{-1.5cm} - \frac{J^{2}}{2J_{\perp}}(\frac{h^{2}}{J_{\perp}^{2}-4h^{2}})
(\xi_{j}^{+}\xi_{j+1}^{+} + {\rm h.c})
- \frac{J^{2}}{8J_{\perp}}(\frac{J_{\perp}^{2} - 2h^{2}}{J_{\perp}^{2} - 4h^{2}})]
\label{end}
\eea
We find that $t_{\perp}$  acts as the strength of a Zeeman-splitting like term
in the LEEH. Bosonising this, we obtain a sine-Gordon Hamiltonian with a 
cosine potential in the dual ($\theta$) field and a magnetic-field term  
\beq 
H=\frac{v}{2}[(\partial_{x}\phi)^{2} + (\partial_{x}\theta)^{2}] 
-\frac{m}{2\pi\alpha}\cos\beta_{1}\theta 
- \frac{\beta_{1} t_{\perp}}{2}\partial_{x}\phi~,
\eeq
where the cosine term arises directly from the BCS-like in 
the effective Hamiltonian, with the bare value of the mass 
$m\sim J^{2}h^{2}/J_{\perp}^{3}$. The velocity, $v$, and sine-Gordon 
correlation exponent, $\beta_{1}$, are 
both functions of the energy scale $J^{2}/J_{\perp}$.
We note that bosonisation of the general XYZ 
Hamiltonian results in the appearance of an additional $4k_{F}$ Umklapp term
~\cite{giamarchi}, $\cos\beta_{2}\phi$, which is irrelevant for a finite 
$t_{\perp}$ and is hence ignored in what follows. When $t_{\perp}$ is 
below a certain critical value, incommensurate Wigner charge order 
(ordering of the $\xi^{z}$) occurs~\cite{giamschulz}. Above this 
critical value, a spin-flop transition orders the system in the 
$x$ direction (i.e., ordering 
of the $\xi^{x}$) via a Kosterlitz-Thouless transition.
For $\beta_{1}^{2}<8\pi$, the cosine in the dual field, $\theta$, is a relevant 
perturbation and orders the dual field. 
The magnetic-field term $\propto t_{\perp}$ leads  
to a ground state with charges which are coherently delocalised on the 
diagonals of each pair of nearest-neighbor rungs; this is an orbital 
antiferromagnet-type ordering with circulating currents in plaquettes
~\cite{tsvelik,giamarchi}.
\par
\subsubsection{LEEH for $J_{\perp} > 0$}
For $J_{\perp}>0$,
and $h>0$, the triplet state
$|+\rangle = |\uparrow\uparrow\rangle$ is the low energy state
on any rung.  For $h=0$, we find that the triplet states
$|+\rangle$ (defined above) and
$|-\rangle = |\downarrow\downarrow\rangle$ are degenerate. Thus,
we can again identify these two states as the subspace which
determines the low-energy physics of the system.
For $h<<J$, we again identify a
pseudospin-1/2 operator $\xi_{j}$ with the low-energy
subspace on each rung, and treat the Hamiltonian $H_{1}$ as a
perturbation (to second order in $J/J_{\perp}$) to obtain the LEEH as
\beq
H = -\frac{J^{2}}{4J_{\perp}}\sum_{j}\xi_{j}^{x}\xi_{j+1}^{x}- 2t\sum_{j}\xi_{j}^{z}~.
\label{2legferro}
\eeq
This is just the 1D TFIM (with ferromagnetic Ising coupling).
In the ordered phase, the ground state has in-chain Wigner CO and dimers on 
every alternate rung.  The disordered phase is a gapped, short-ranged charge-dimer liquid.  At $t=J^{2}/4J_{\perp}$, the quantum critical point describes a gapless charge-dimer liquid with $\omega/T,vk/T$ QC scaling, exactly as was
described before.  Transposing the results obtained before, we conclude that the dc resistivity, optical conductivity, electronic Raman and dielectric responses
will be exactly described by the same scaling functions 
(eqs.(\ref{chiasym})-(\ref{ramanshape})) with the 
gap, $\Delta_{\tau}$,  now being the CO gap of the ladder problem 
($H$ in eq.(\ref{hcomplete})).
Very interestingly, exactly such behavior is observed in undoped ladder system
$Sr_{14}Cu_{24}O_{41}$~\cite{blum} and attributed to a longitudinal, collective charge fluctuation mode, exactly as described here.
\subsection{$|t_{\perp}| >> |J|,~ J_{\perp}$}
For the case of $|J_{\perp}| >> |J|,~ t_{\perp}$, we can again treat the 2-chain 
system as that composed of strongly coupled rungs which are weakly coupled to their 
neighbours. Thus, we treat $J$ as a 
weak perturbation on the zeroth-order system of rungs defined by the large
coupling $J_{\perp}$, giving $H_{eff} = H_{0} + H_{1}$ where
\bea
H_{0} &=& -h\sum_{j,a} \tau_{j,a}^{z}
- \frac{t_{\perp}}{2}\hspace*{-0.2cm}\sum_{j,a,b\neq a}\hspace*{-0.2cm}(\tau_{j,a}^{+}\tau_{j,b}^{-}
+ {\rm h.c})\nonumber\\*
H_{1} &=& - J\sum_{j,a} \tau_{j,a}^{x}\tau_{j+1,a}^{x}
- J_{\perp}\sum_{j,a,b\neq a}\tau_{j,a}^{z}\tau_{j,b}^{z}
\label{start1}
\eea
where the effective magnetic field is given by $h=2t>0$.
\par
For $t_{\perp} >0$ and
$h = t_{\perp}/2$, we find that the triplet zero state
$|+\rangle = \frac{1}{\sqrt{2}}(|\uparrow\downarrow\rangle +
|\downarrow\uparrow\rangle)$ and the triplet up state $|-\rangle = 
|\uparrow\uparrow\rangle$ are degenerate on any rung and 
are separated from 
all other states by a large gap of order $t_{\perp}$. Thus, these two states 
define the subspace which will determine the low-energy physics
of the system.
Identifying a pseudospin-1/2 operator $\xi_{j}$ with the low-energy
subspace on each rung, we treat the Hamiltonian $H_{1}$ as a
perturbation to obtain 
the LEEH as 
\bea
H &=& \sum_{j}[-J_{\perp}\xi_{j}^{z} + \frac{J}{2}(\xi_{j}^{+}\xi_{j+1}^{-} 
+ {\rm h.c})]
\label{end1}
\eea 
Note that, unlike the LEEHs derived earlier, this LEEH is at first order in 
$J/t_{\perp}$ and $\tJp/t_{\perp}$. Further, we have checked that the terms 
obtained at next order in the perturbative expansion are considerably smaller 
and do not introduce anything qualitatively new; it is, therefore, sufficient 
to stop at this order. 
\par
The expression (\ref{end1}) is the Hamiltonian for 
the isotropic XY model in a transverse magnetic field. This theory is, again, 
exactly solvable and has a $T=0$ QCP in its phase diagram at 
$J_{\perp}^{c}=|J|/2$. Further, there exists an equivalence between the classical 
2D Ising model at finite $T$ and the $T=0$ isotropic quantum XY model in a 
transverse field~\cite{suzuki}: the high (low) $T$ regions of the former 
with $T>T_{c}$ ($T<T_{c}$) map onto the high (low) $J$ regions of the latter 
with $J_{\perp}>J_{\perp}^{c}$ ($J_{\perp}<J_{\perp}^{c}$). The critical 
exponents associated with the finite-$T$ thermal phase transition in the 
classical 2D Ising model are also identical to those associated with the 
$T=0$ quantum phase transition in the XY model in a transverse field. Thus, 
we can expect scaling forms for the various response functions of the 
system in the quantum critical region of the phase diagram (lying just 
above the QCP in the $T$ direction), as discussed earlier. In the ordered 
phase, the ground state of the system has either rung-dimer or rung-diagonal 
dimer order, while in the disordered state is characterised again by a gapped 
charge-dimer liquid. 
\subsection{LEEH for hole-doped ladder} 
Upon doping the ladder with holes,
while a single hole experiences a linear confining potential in the
Wigner (Ising-like) or Peierls (dimerized) CO background, a pair of
holes on the same rung is free to propagate. One can then describe the
hole-pair as a hard-core boson, representing its creation and annihilation
operators using the spin-1/2 operators $\sigma^{\pm}$; the local charge
density is then described by $\sigma^{z}$. Following~\cite{tsvelik}, we
find the
LEEH describing the dynamics of such hole-pairs to be the XXZ model
in an external magnetic field
\beq
H=\sum_{j}[-\frac{t_{\mathrm{h}}}{2}(\sigma_{j}^{+}\sigma_{j+1}^{-} + {\rm h.c})
- u_{\mathrm{h}}\sigma_{j}^{z}\sigma_{j+1}^{z} - \mu\sigma_{j}^{z}]
\eeq
where $t_{\mathrm{h}} \sim \tJ^{2}/\tJp$ is the pair-hopping matrix element,
$u_{\mathrm{h}}$ is the Coulomb interaction between pairs on nearest-neighbour
rungs and $\mu$ is the chemical potential of the holes. The phase
diagram of this model is known~\cite{tsvelik}: for $\mu=0$ and
$u_{\mathrm{h}}>t_{\mathrm{h}}$, the ground state is an insulating CDW of hole pairs.
Beyond a critical $\mu_{\mathrm{c}}=f(u_{\mathrm{h}},t_{\mathrm{h}})$, the system has a ground
state described by Bose condensation of hole pairs. In fact, from
the bosonisation analysis of the equivalent $S=1/2$ XXZ model in an
external Zeeman field, we know that
$<\sigma_{i}^{z}\sigma_{i+r}^{z}> \simeq r^{-1/\alpha}$ and
$<\sigma_{i}^{+}\sigma_{i+r}^{-}> \simeq r^{-\alpha}$ where
$\alpha=1/2-\pi^{-1}\sin^{-1}(2u_{\mathrm{h}}/t_{\mathrm{h}})$. Clearly, for $\alpha<1$,
the ground state has dominant superconducting correlations. This is
true for {\it both} the cases described above: in the first case, we
have a Bose condensate of intrachain pairs of holes, while in the
second hole pairs on individual rungs Bose condense, describing two
possible superconducting types in the ladder system. This finding
matches our conclusions obtained from a weak coupling analysis
~\cite{next}, and
thus constitutes a generic feature of undoped/doped strongly correlated
ladder systems.
\section{Dimensional Crossover in Coupled Chain and Ladder Systems} 
Having studied the case of the single chain at strong-coupling as well 
as those of strongly coupled 2-chain ladder systems of such chains, we 
now turn our attention to the invesigation of the case of when many such 
chains and ladders are coupled to one another through transverse couplings 
of the kind studied in the previous section. We have recently studied~\cite{next} 
in detail the consequences of transverse couplings between many TFIM systems 
using the random phase appromximation (RPA)~\cite{scalapino,schulz,carr}. 
In what follows, therefore, we will rely largely on the derivations and 
results give there~\cite{next}. With the transverse field Ising model (TFIM) having 
appeared as an effective theory for both the case of a single electronic chain 
at strong coupling as well as proving to be the LEEH for a strongly coupled 2-leg 
TFIM ladder system, we will focus first on dealing with the case of TFIM chains 
and ladders. We then pass to a study of some of the other ladder LEEH theories 
for another work.
\par 
Let us begin by setting out what we can expect to happen generically upon 
coupling many TFIM chain or ladder systems. In the absence of any transverse 
couplings, as discussed earlier, by defining the coupling 
$g=|(V-J/4-2P-2t)/(V-J/4-2P)|$, the phase diagram at $T=0$ is simple (see 
Fig.(\ref{tfimphasediag})), 
with an ordered phase for $2t<V-J/4-2P$ ($g<0$), a quantum disordered phase 
for $2t>V-J/4-2P$ ($g>0$) and a quantum critical point at $2t=V-J/4-2P$ ($g=0$). A 
finite transverse coupling, denoted by ${\cal O}_{\perp}$, causes 
the ordered phase to be extended to finite temperatures (with a critical 
temperature $T_{c}$ for the case of $2t=V-J/4-2P$) with a first order 
phase boundary ending at a new quantum critical point (QCP) $g_{c}=\Delta_{c}/V$ at 
$T=0$~\cite{carr}. As for the simple TFIM~\cite{sachdev}, there exists a 
``quantum critical" region just above the QCP and to the right of the ordered 
phase, with a crossover to the disordered phase at finite $T$. This is shown 
schematically in the $T-g$ phase diagram given below in Fig.(\ref{qcp}). The 
transition belongs to the 3D Ising universality class while the QCP to the 
4D Ising universality class~\cite{itzdrouffe}.
\begin{figure}[htb]
\begin{center}
\scalebox{0.35}{
\psfrag{1}[bl][bl][3.5][0]{$T_{c}$}
\psfrag{2}[bl][bl][4]{$T$}
\psfrag{3}[bl][bl][3.5][0]{$0$}
\psfrag{4}[bl][bl][3.5][0]{$g_{c}$~({\bf QCP})}
\psfrag{5}[bl][bl][4][0]{$g$}
\psfrag{6}[bl][bl][3][0]{Ordered}
\psfrag{7}[bl][bl][3][0]{Gapless}
\psfrag{8}[bl][bl][3][0]{Disordered}
\psfrag{9}[bl][bl][3][0]{Phase}
\psfrag{10}[bl][bl][3][0]{Phase}
\psfrag{11}[bl][bl][3][0]{Critical}
\psfrag{12}[bl][bl][3][0]{Region}
\psfrag{13}[bl][bl][3][0]{Quantum}
\includegraphics{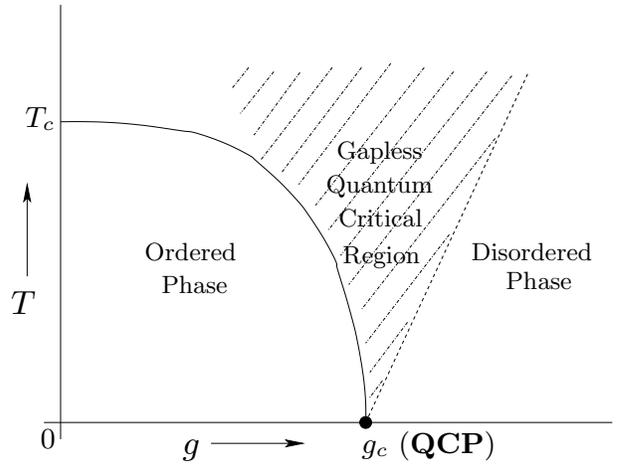}
}
\end{center}
\caption{The $T-g$ phase diagram for the case of many TFIM systems coupled 
by a transverse coupling $t_{\perp}$ (or $J_{\perp}$). The $T=0$ Ordered 
Phase of the uncoupled TFIM is now extended, with a phase boundary which 
has a value $T_{c}$ for the model at $g=0$ and a $T=0$ quantum critical 
point (QCP) at $g\equiv g_{c}$. The hashed region immediately to the right 
of the ordered phase and just above the QCP is the gapless quantum critical 
region (described in the text). The dashed line represents a finite $T$ 
crossover from the quantum critical region to a Disordered phase.} 
\label{qcp}
\end{figure}
\par
In keeping with the fact that the passage to higher dimensions is essentially 
a nonperturbative phenomenon ~\cite{giamarchi}, we use the RPA method to 
study the physics of dimensional crossover. Put another way, we are interested 
in gaining an understanding of how lower dimensional systems (here, 
our one-dimensional strongly correlated chains), when coupled to one another, 
go from being nearly isolated to a anisotropic strongly coupled system in 
higher dimensions. It is worth noting that, while the mean-field-like approach 
of RPA is exact only in infinite dimensions (i.e., infinite coordination 
number), its application to the physics of coupled quasi-1D spin systems 
for small coordination numbers (i.e., lower dimensions) has met with considerable 
success~\cite{irkhin,bocquet2}. Further, while a naive mean-field 
treatment of single particle-hopping between fermionic chains is not possible 
as a single fermion operator has no well-defined classical limit
~\cite{giamarchi}, we are able to treat two-particle hopping processes between 
our underlying chain and ladder systems via RPA by working with an effective 
theory in terms of pseudospins.
This is justified because the presence of the large on-site Hubbard coupling 
in our underlying model makes the single particle hopping irrelevant (in a 
RG sense), while two-particle processes (including hopping terms) can be 
crucial in determining the phase diagram~\cite{giamarchi}. A full treatment 
including single-particle hopping will require a treatment using 
chain-dynamical mean field theory (c-DMFT)~\cite{arrigoni} and will be the 
focus of a future work. 
\par
For a transverse coupling ${\cal O}_{\perp}$, the RPA method involves computing 
the dynamical spin susceptibility $\chi$ 
of the coupled system in the disordered phase as 
\beq
\chi (\omega, k, \vec{k}_{\perp}) = 
[\chi_{1D}^{-1}(\omega,k)-{\cal O}_{\perp}(\vec{k}_{\perp})]^{-1} 
\label{rpamain}
\eeq
in terms of the frequency $\omega$, the longitudinal and transverse wave-vectors 
$k$ and $\vec{k}_{\perp}$ respectively. $\chi_{1D}$ is the dynamical spin 
susceptibility of a single TFIM system, to be calculated assuming incipient 
order along the $\tau^{x}$ direction in pseudospin space 
\beq
\chi_{1D}(\omega,k) = -i\sum_{n}\int_{0}^{\infty}dt e^{i(\omega t - kn)}
\langle\lbrack\tau^{x}(t,n),\tau^{x}(0,0)\rbrack\rangle
\label{1dchi}
\eeq
and the transverse coupling ${\cal O}_{\perp}(\vec{k}_{\perp})
\sim z_{\perp}{\cal O}_{\perp}(\vec{k}_{\perp}=0)$, for each TFIM system having 
a coordination number of $z_{\perp}$. Then, a divergence in the 
dynamical pseudospin correlation function $\chi (\omega,k,\vec{k_{\perp}})$ 
signifies an instability towards the formation of an ordered state. In this 
way, it is possible to compute the quantities $g_{c}$ at $T=0$, $T_{c}$ for 
the case of $g\equiv\Delta/(V-J/4-2P)=0$ (single pseudospin chain mass 
$\Delta=|V-J/4-2P-2t|$ for the TFIM), 
the shape of the phase boundary near the QCP, the dynamical spin susceptibility 
$\chi(\omega,k,\vec{k}_{\perp})$ at the QCP as well as the dispersion in the 
transverse directions for small ${\cal O}_{\perp}$ and close to the QCP~\cite{carr}. 
We focus our attention mainly on, and in the neighbourhood of, the QCP in order to 
assess the role played by the critical quantum fluctuations in determining 
the physics of deconfinement and dimensional crossover in our system. In 
this, we will often be aided by the integrability and conformal invariance of the 
unedrlying pseudospin model (e.g., for the TFIM, at $2t=V-J/4-2P$,~${\cal O}_{\perp}=0$
~\cite{lieb,sachdev}). We can then exploit the nonperturbative results for a 
single pseudospin system (as long as the mass scale $\Delta<<1$), while dealing 
with the physics of the transverse couplings at a mean-field level.
\par
We learn from the results presented below that, in passing from the 
interior of the ordered phase towards the phase boundary, the excitation gaps 
decrease together with a gradual growth of the dispersion in the transverse 
directions. The spectrum is gapless at the QCP. At 
low $T$, directly in the region above the QCP, the spectrum and dynamics are 
mainly governed by the QCP while the thermal excitations are described by the 
associated continuum quantum field theory~\cite{sachdev}. The dimensional 
crossover is thus characterised 
by the growth of the dispersion in the transverse directions close to the QCP, 
while the $T=0$ deconfinement transition is characterised by the vanishing of 
all mass gaps at the QCP. 
\par
We can, thus, begin with the case of transverse 
couplings $t_{\perp}$ and $J_{\perp}$ in a system of many TFIM chains 
(as given in equn.(\ref{twochain})). As 
a detailed study of a generic situation of coupled TFIM systems has been carried 
out in Ref.(\cite{carr,next}), it is sufficient to quote the main results here 
for the case of $t_{\perp}$. As discussed below, this is done since the results 
for the $J_{\perp}$ coupling are found to be qualitatively similar. The 
critical temperature $T_{c}$ for 
the case of $g=0$ is obtained from the $\chi_{1D}(\omega=0,k=0)$ at finite $T$
\beq
\chi_{1D}(\omega=\hspace*{-0.1cm}0\hspace*{-0.1cm}=k)\hspace*{-0.1cm}=
\frac{c_{2}}{(V-J/4-2P)}(\frac{2\pi T}{(V-J/4-2P)})^{-7/4}
\label{omega0k01dchi}
\eeq 
where the constant $c_{2}=\sin(\pi/8){\rm B}^{2}(1/16,7/8)$ and ${\rm B}(x,y)$ 
is the Euler Beta function. Thus, we find, from the condition for the divergence 
of the susceptibility $\chi$
\beq
\frac{T_{c}}{(V-J/4-2P)} = c_{3}(\frac{z_{\perp}t_{\perp}}{(V-J/4-2P)})^{4/7}
\label{critT}
\eeq
where the constant $c_{3}= c_{2}/(2\pi) = 2.12$.
\par
We next compute the critical value of the transverse coupling 
$g_{c}$ at $T=0$. For this, we can use the expression for slightly 
off-critical $\chi_{1D}$ for the case when the mass of the 
single TFIM system is very small ($m=\Delta_{c} << 1$). This, for small 
$\omega$, is given by   
\beq
\chi_{1D}(\omega,k) \simeq \frac{Z_{0} V (\Delta_{c}/V)^{1/4}}
{\omega^{2}-(v k)^{2} - \Delta_{c}^{2}}
\label{offcrit1dchi}
\eeq
where the velocity $v=(V-J/4-2P)\alpha$ (where $\alpha$ is the lattice spacing) and 
$Z_{0}=1.8437$. Then, from the divergence of the susceptibility of the coupled 
system, $\chi(\omega,k,\vec{k}_{\perp})$ 
\beq
\chi_{1D}^{-1}(\omega,k) = z_{\perp}t_{\perp}(\vec{k}_{\perp}=0),
\label{diverge}
\eeq 
we get, for the case of $\omega=0=k$,
\beq
\frac{(V-J/4-2P)}{Z_{0}} g_{c}^{7/4}\approx z_{\perp}t_{\perp}
\eeq
where we've dropped the argument of $\vec{k}_{\perp}=0$ in $t_{\perp}$ 
for the sake of convenience. Thus, we get
\beq
g_{c}\approx c_{1}(\frac{z_{\perp}t_{\perp}}{(V-J/4-2P)})^{4/7}
\label{critg}
\eeq
where the constant $c_{1}=Z_{0}^{4/7}=1.42$. Precisely the same expression for the 
mass in the ordered phase and very close to the QCP, $\Delta=g_{c}(V-J/4-2P)$, is 
also obtained 
by carrying out a self-consistent treatment of the effective magnetic field, 
$h=z_{\perp}t_{\perp}\langle\tau^{x}\rangle$ in the TFIM for a single chain.
For this, one uses the slightly off-critical susceptibility 
for the 1D TFIM given earlier (eqn.(\ref{offcrit1dchi})) with the mass 
$\Delta$ replaced by $\Delta(1+(h/(V-J/4-2P))^{2})$~~\cite{delfino}. From this result,
the authors of Ref.(\cite{carr}) concluded that  the dispersion in the transverse 
directions in the ordered phase and close to the QCP is much stronger than that 
deep in the ordered phase. 
\par
The susceptibility $\chi$ for the coupled system at the QCP for small 
$\vec{k}_{\perp}$ can also be computed by using the relation for the 
slightly off-critical 
$\chi_{1D}$ given earlier, eqn.(\ref{offcrit1dchi}), together with the relation 
$\Delta_{c}^{2}=g_{c}^{1/4}(V-J/4-2P)t_{\perp}$ in eqn.(\ref{rpamain}). This leads to 
\beq
\chi (\omega,k,\vec{k}_{\perp})\sim\frac{Z_{O} (V-J/4-2P) g_{c}^{1/4}}
{\omega^{2} - (v k)^{2} - (\vec{v}_{\perp}\cdot\vec{k}_{\perp}){2}},
\label{critchi}
\eeq
where $|\vec{v}_{\perp}|^{2}= (Z_{0} (V-J/4-2P) g_{c}^{1/4}/2)
d^{2}t_{\perp}(\vec{k}_{\perp}=0)/d\vec{k}_{\perp}^{2}$ is gained 
by a Taylor expansion to second order~\cite{carr,next}. The shape 
of the phase boundary at low $T$ can now be determined by using the 
$\chi_{1D}$ of the TFIM at low $T$~~\cite{sachdev}
\beq
\chi_{1D} (\omega, k) = \frac{Z_{0}(\alpha\Delta/v)^{1/4}}
{(\omega + i/\tau_{\psi})^{2} - (v k)^{2} - \Delta^{2}}, 
\label{offcritfiniteTchi}
\eeq
where $\tau_{\psi}=\frac{\pi}{2T}e^{\Delta/T}$ is the dephasing time 
due to quantum fluctuations. Then, for $\omega=0=k$ in $\chi_{1D}$, 
the eqn.(\ref{diverge}) gives
\beq
\ln T - \frac{\Delta}{T} = \ln m + \ln \Lambda,
\label{transcen}
\eeq
where $\Lambda=\frac{\pi}{2}(\frac{Z_{0}t_{\perp}}{g^{7/4}V} - 1)^{1/2}$.
The expression (\ref{transcen}) given above has an approximate solution~\cite{carr,next}
\beq
T_{c} = \frac{\Delta}{\ln(1/\Lambda) - \ln\ln(1/\Lambda)}~.
\label{phaseboundary}
\eeq
This relation gives us the shape of the phase boundary for low $T$ and close to 
the QCP. We have, therefore, derived the important features of the $T-g$ phase diagram 
for the case of the $t_{\perp}$ transverse coupling (Fig.(\ref{qcp})) given above. 
A RPA calculation for the other transverse coupling, 
$J_{\perp}$, for critical TFIM chains (i.e., the Hamiltonian (\ref{twochain}) with 
$V-J/4-2P=2t$, $U_{\perp}=J_{\perp}$ and $t_{\perp}=0$) can also be carried 
out~\cite{carr}. This is guaranteed by the integrability of this 
Hamiltonian~\cite{zamol} and the fact that it falls into the same universality as 
the TFIM. The RPA calculation thus leads to an similar set of 
relations for $g_{c}$, $T_{c}$, 
the susceptibility $\chi$, dispersion in the transverse directions as 
well as the shape of the phase boundary close to the QCP to those obtained 
earlier, but with $t_{\perp}$ replaced by $J_{\perp}$ everywhere. In this way, 
we obtain essentially the 
same $T-g$ phase diagram for the case of the $J_{\perp}$ transverse coupling; 
the only notable difference is that the spectrum of the ordered phase is now 
obtained from the exact solution of the 1D TFIM in a longitudinal 
field~\cite{zamol}. 
\par
We now turn to a discussion of the results of a RPA treatment for the case of 
coupled 2-leg TFIM ladders, by using the LEEHs obtained in various limits in 
section III. For the case of the strongly coupled 2-leg Ladder LEEH with the coupling 
$J_{\perp}>0$ (equ.(\ref{2legferro})), as we again find the TFIM Hamiltonian as 
the effective theory, we can safely conclude that an RPA treatment for many 
such coupled ladders will lead to the same results as those given above and a 
phase diagram identical to Fig.(\ref{qcp}). A RPA treatment of the LEEH derived 
for the case of $J_{\perp}<0$ (equn.(\ref{end})) can be carried out easily for some 
special cases: (i)$h=0=t_{\perp}$, (ii) $t_{\perp}=(J^{2}/J_{\perp})
(J_{\perp}^{2}-2h^{2})/(J_{\perp}^{2}-4h^{2})$ and (iii) $J_{\perp}^{2}=2h^{2}$. We
discuss each of these in turn. 
\par
For $t_{\perp}=0=h$, 
the LEEH for $J_{\perp}<0$ reduces simply to that of the 1D Ising model with an 
exchange coupling $-J^{2}/2J_{\perp}$. We can now treat interchain coupling terms 
$\bar{t}(\xi^{x}_{j,a}\xi^{x}_{j,b} + \xi^{y}_{j,a}\xi^{y}_{j,b})$ and 
$\bar{U}\xi^{z}_{j,a}\xi^{z}_{j,b}$ (where $a,b$ are chain indices, 
$\bar{t}\sim J^{4}/J_{\perp}^{3}$ and $\bar{U}\sim J^{3}/J_{\perp}^{2}$) in turn 
via RPA. In this program, we can assume order along a given direction in pseudospin 
space and then replace the appropriate interchain coupling term by an effective 
field term (in the spirit of a mean-field treatment) and solve the Hamiltonian in a 
self-consistent manner. Thus, we can see that for the $\bar{t}$ spin flip term, 
such a mean-field treatment assumes order along $\xi^{x}$, say, and thus
has an effective magnetic field $h=z_{\perp}\bar{t}\langle\xi^{x}\rangle$ (where 
$z_{\perp}$ is the coordination number of any ladder system). This leads to the
 self-consistent effective Hamiltonian of the coupled ladder problem again taking 
the form of the 1D TFIM. Computing the critical value of the gap at the QCP in a 
self-consistent manner (as discussed earlier) gives
$g_{c} = J^{2}/(2J_{\perp})(2z_{\perp}\bar{t}J_{\perp}/J^{2})^{4/7}$. In this way,
it is clear that the self-consistent 
solution of this effective theory will again give rise to a phase diagram 
like Fig.(\ref{qcp}). 
\par
An identical calculation for the $\bar{U}$ Ising transverse coupling term 
can be carried out for the case of $t_{\perp}=(J^{2}/J_{\perp})
(J_{\perp}^{2}-2h^{2})/(J_{\perp}^{2}-4h^{2})$. This is the case of the critical 
TFIM in a longitudinal field, which is integrable and belongs to the same universality 
class as the TFIM~\cite{zamol}. The RPA calculation assumes 
order along $\xi^{z}$ and has an effective (self-consistent) magnetic field 
$h=z_{\perp}\bar{U}\langle\xi^{z}\rangle$ in a 1D Ising model in a longitudinal 
field. Using the exact solution of the problem~\cite{zamol}, together 
with the divergence condition in the RPA, gives us the critical gap as 
$g_{c} = J^{2}/(2J_{\perp})(2z_{\perp}\bar{U}J_{\perp}/J^{2})^{4/7}$. Again, we reach 
qualitatively similar conclusions with regards to the phase diagram for this 
transverse coupling. Finally, we note that the LEEHs for both the cases of 
$J_{\perp}<0, J^{2}=2h^{2}$ and $t_{\perp}>>J,J_{\perp}$ are those of the 
XY model in a transverse magnetic field; a detailed RPA investigation of this problem 
will be presented elsewhere. 
\section{Comparison with recent numerical works}
We present here a discussion of the relevance of our work to some numerical 
investigations that have been carried out on strongly correlated single chain,
2-leg ladder systems as well as coupled TFIM systems. We begin with a discussion of 
the early work of Capponi et al.~\cite{capponi}. The study assessed the effects 
of long-range Coulomb interactions on the phase diagram of a 
finite-size one-dimensional system of spinless electrons at $1/2$-filling. 
The authors concluded 
from an exact diagonalisation analysis that for intermediate strengths, the 
presence of extended range interactions caused an enhancement of the metallic 
nature of the system; this is in contrast with the fact that in the thermodynamic 
limit, such a system would always be driven by the logarithmic divergence of the 
long wavelength part of the interactions towards insulating character. While the 
metallic phase was observed to have a vanishing gap, it did not agree with the
predictions of conformal field theory. Increasing the strength of the Coulomb 
interactions caused a crossover towards a localised $2k_{F}$ CDW phase. Note that 
a gapless metallic phase arising from a Wigner CO phase as the strength of 
the Coulomb interactions is reduced is not in contradiction with our 
results for the charge sector of 
the single chain in the strong-coupling regime: the effective TFIM 
model at $T=0$ has a gapless QCP emerging from a CO phase as the nearest neighbour 
hopping $t$ grows to a critical value. Capponi et al., however, found no 
signatures of the Peierls CO phase for the single chain discussed in section II.  
\par
There are a few notable works on quasi-1D strongly correlated models at 
$1/4$-filling with extended interactions. Riera et al.~\cite{riera} studied the case of 
$1/4$-filled chain/ladder Hubbard and $t-J$ systems, but which also include Holstein 
and/or Peierls-type couplings to the underlying lattice. Their findings reveal 
coexisting charge and spin orders in both chain and ladder systems. Specifically, 
for the case of their chain system, by keeping only on-site and nn 
repulsion together with an on-site Holstein-type coupling of the electronic 
density to a phonon field, their phase diagram (Figs.4) reveal separate phases 
with Wigner as well Peierls-type charge order. This is in keeping with our 
findings, but the origin of the Peierls order in the two cases are different: 
in our work, it originates from the competition of the nnn coupling $V_{2}$ with 
the nn coupling $V_{1}$, while in their work, it needs the Holstein coupling to 
the lattice. Riera et al. find a similar phase diagram (Figs.5) for the case 
of an extended $t-J$ model (i.e., including nnn t and J couplings) with a  
Holstein coupling. The addition of a Peierls-type coupling leads to a 
spin-Peierls instability, i.e., the formation of a dimerised spin order, 
which coexists with the Peierls-type charge order. Again, while this matches our 
findings, the origins are different. Qualitatively similar conclusions are also 
reached by the authors in their study of an anisotropic 2-leg $t-J$ ladder with 
an extended on-chain nn coupling and Holstein/Peierls-type lattice couplings 
(Figs.2 and 3).
\par
Vojta et al.~\cite{vojta} studied the problem of a 
strongly correlated electronic problem at $1/4$-filling and with extended 
Hubbard interactions (keeping only a nn repulsion) using the 
DMRG method. The phase diagram they obtained contains several phases with 
charge and/or spin excitation gaps. While a comparison of our work with this 
study is hindered by the fact that the DMRG analysis does not have the crucial 
element of the nnn repulsion ($V_{2}$ in our work), Fig.2 of 
that work reveals that for the case of $U>>V_{1}>t$, the authors indeed find a 
charge-ordered CDW state (i.e, the Wigner charge ordered state of 
discussed in section II) with an excitation gap in the spin sector as well. This is 
in conformity with our finding of a Wigner charge-ordered state with a spin 
gap for the case of the $U_{\rho}$ coupling being the most relevant under RG. 
Further, the $t_{\perp}$ of that study corresponds to the single-particle 
hopping between the legs while our work has focused on the effects of 
two-particle hopping. Finally, with the on-site Hubbard coupling, $U$, being 
the largest in the problem, we are unable to see any phase-separated state 
in our phase diagram (as observed by Vojta et al.). 
\par
We end by commenting on a very recent DMRG studies of Konik et al.~\cite{konik} 
on coupled TFIM systems in an effort at studying two dimensional coupled arrays 
of one dimensional systems. By starting with a reliable spectrum truncation 
procedure for a single TFIM chain (which relies on the underlying continuum 
1D theory being either conformally invariant or gapped but integrable), the 
authors then implement an improvement of their DMRG algorithm using first-order 
perturbative RG arguments. Their results for a $J_{\perp}$ coupling of the 
TFIM chains confirms the accuracy of the RPA analysis of Ref.(\cite{carr}) 
and the present work in computing quantities like the single chain excitation 
gap (which is found to vanish at a critical $J_{\perp}$) and dispersion of 
excitations in the coupled system as a function of $J_{\perp}$. Their results 
indicate that the RPA method and the DMRG analysis agree very well upto values 
of $J_{\perp}$ of the order of 
the gap. This method also appears to give accurate values of critical 
exponents related to the ordering transition. Thus, this numerical approach 
appears to provide a confirmation of the interplay of the QCP in 
the TFIM and the transverse coupling in driving the dimensional crossover and 
deconfinement transition. Such an approach, therefore, holds much promise for 
the numerical investigations of such phenomena in systems with similar 
ingredients.
\section{Conclusions}
To conclude, we have explored the strong-coupling limit of 
strongly correlated $1/4$-filled single chain and two-leg ladder models using 
a variety of methods. The charge sector of a $1/4$-filled 1D model of 
electrons with extended 
interactions is, in the regime of the on-site Hubbard term being the largest 
and the hopping strength the smallest, found to lead to an effective theory 
of the 1D transverse field Ising model. The ordered 
phases in the charge sector are found to belong to either the Wigner or 
Peierls type CO. The two kinds of CO are then found to give rise to a 
gapless AF ground state and a dimerised state respectively in the spin 
sector. The integrability of the 1D TFIM 
allows us to make considerable progress in computing various thermodynamic 
quantities (e.g., response functions), especially in the quantum critical 
region lying just above the $T=0$ quantum critical point (QCP). 
\par
Strongly 
coupled 2-leg ladder systems composed of such chains are also studied in 
various limits, with the low energy effective theory generically being found 
to be described once again by an exactly solvable 1D model with a QCP. The 
varying of interchain couplings in the 2-leg ladder is found to give rise 
to a variety of charge ordered phases (e.g., in chain Wigner CO, rung-dimer 
as well as orbital antiferromagnet type charge order). This is in conformity 
with our recent studies on the weak coupling phase diagram for a coupled 2-leg TFIM 
system~\cite{next}. Doping such ladders with holes is also found to give rise 
to superconductivity. RPA studies on the effects of transverse couplings 
connecting many such chain and ladder systems presented a generic phase diagram 
for the coupled system containing an ordered phase extending to finite $T$ and 
with a phase boundary ending in a quantum critical point. These 
calculations also stressed the importance of the role of the QCP in the mechanism 
responsible for the dimensional crossover in the quantum critical region lying 
at finite $T$ just above the QCP and the accompanying $T=0$ deconfinement transition. 
\par
Significantly, the favourable comparison of our findings, as discussed earlier, 
for the strongly coupled single chain and 2-leg ladder systems  
with experimentally observed phenomena in prototype examples like organics 
(TMTSF) and $Sr_{14}Cu_{24}O_{41}$ respectively strongly suggests that these 
systems lie in close proximity to an underlying QCP associated with charge order. 
This constitutes a significant advance in our understanding of the physical 
responses of these systems in a new theoretical framework. Further, the 
robustness of the 
dimensional crossover mechanism found in this work leads us to conclude that
critical quantum fluctuations associated with a QCP can quite generically  
enhance the dispersion in the transverse dimensions for anisotropic systems, i.e., 
facilitate the passage from the gapped phases of lower dimensional systems 
to the gapless phases of the coupled system in higher dimensions. 

\begin{acknowledgement}
We thank E. M\"{u}ller-Hartmann, G. I. Japaridze, 
F. Franchini, L. Dall'Asta, 
S. Basu, S. T. Carr, A. Nersesyan and M. Fabrizio for several fruitful discussions.
\end{acknowledgement}



\begin{thebibliography}{}



\bibitem{mott} A classic text on the subject is N. F. Mott, {\it Metal-Insulator 
Transitions} (Taylor and Francis, London, 1990).

\bibitem{imada} M. Imada, A. Fujimori and Y. Tokura, Rev. Mod. Phys. 
{\bf 70}, (1988) 1039.

\bibitem{org} D. Jerome, {\it Organic Superconductors: From $(TMTSF)_2PF_6$ to 
Fullerenes} (Marcel Dekker, New York, 1994), pgs. 405.

\bibitem{salamon} M. B. Salamon and M. Jaime, Rev. Mod. Phys. {\bf 73}, 
(2001) 583.

\bibitem{mccaron} E. M. McCarron III {\it et al.}, Mater. Res. Bull. {\bf 23}, 
(1988) 1355; M. Uehara {\it et al.}, J. Phys. Soc. Jpn. {\bf 65} (1996) 2764.


\bibitem{ashcroft} N. W. Aschcroft and N. D. Mermin, Solid State Physics (Holt, 
Rinehart and Winston, New York, 1976). 

\bibitem{giamarchi} Thierry Giamarchi, {\it Quantum Physics in One Dimension} 
(Oxford Univ. Press, Oxford, 2004) and references therein.

\bibitem{capponi} S. Capponi, D. Poilblanc and T. Giamarchi, Phys. Rev. B 
{\bf 61}, (2000) 13410.

\bibitem{tsuchiizu} M. Tsuchiizu and A. Furusaki,  Phys. Rev. B {\bf 69}, 
(2004) 035103 and references therein.

\bibitem{yoshioka} H. Yoshioka, M. Tsuchiizu and Y. Suzumura, J. Phys. Soc. 
Jpn. {\bf 69} (2000) 651.

\bibitem{tsuorig} M. Tsuchiizu and E. Orignac, J. Phys. Chem. Solids {\bf 63} 
(2002) 1459.

\bibitem{yoshi1} H. Yoshioka, M. Tsuchiizu and H. Seo, J. Phys. Soc. 
Jpn {\bf 75}, (2006) 063706.

\bibitem{riera} J. Riera and D. Poilblanc, Phys. Rev. B {\bf 59}, (1999) 2667.

\bibitem{kuwabara} M. Kuwabara, H. Seo and M. Ogata, J. Phys. Soc. Jpn. 
{\bf 72}, (2003) 225.

\bibitem{yoshi2} H. Yoshioka, M. Tsuchiizu and H. Seo, cond-mat/0708.0910

\bibitem{hubbard} J. Hubbard, Phys. Rev. B {\bf 17}, (1978) 494.

\bibitem{scalapino} D. J. Scalapino, Y. Imry and P. Pincus, Phys. Rev. B 
{\bf 11}, (1975) 2042.

\bibitem{schulz} H. J. Schulz, Phys. Rev. Lett. {\bf 77}, (1996) 2790.

\bibitem{carr} S. T. Carr and A. M. Tsvelik, Phys. Rev. Lett. {\bf 90}, 
(2003) 177206.

\bibitem{emery} V. J. Emery and C. Noguera, Phys. Rev. Lett. {\bf 60}, (1988) 631. 

\bibitem{kats} A. K. Zhuravlev and M. I. Katsnelson, Phys. Rev. B {\bf 64}, 
(2001) 033102. 

\bibitem{peschel} I. Peschel, Z. Phys. B {\bf 45}, (1982) 339.

\bibitem{liebmann} R. Liebmann, {\it Statistical Mechanics of Periodic 
Frustrated Ising Systems}, Lecture Notes in Physics Vol.~{\bf 251} 
(Springer-Verlag, Berlin, 1986).

\bibitem{lieb} E. Lieb, T. Schultz and D. Mattis, Ann. Phys.(N.Y.) 
{\bf 16}, (1961) 406.

\bibitem{niemeijer} Th. Niemeijer, Physica {\bf 36}, (1967) 377; 
{\it ibid.}, Physica {\bf 39}, (1968) 313.

\bibitem{chakrabarti} B. K. Chakrabarti, A. Dutta and P. Sen, {\it Quantum 
Ising Phases and Transitions in Transverse Ising Models}, Lecture Notes in 
Physics Vol. m 41 (Springer-Verlag, Berlin, 1996).

\bibitem{sachdev} S. Sachdev, {\it Quantum Phase Transitions} (Cambridge 
Univ. Press, Cambridge, 1999) and references therein.

\bibitem{kopp} A. Kopp and S. Chakravarty, Nature Physics {\bf 1}, (2005) 53.


\bibitem{tsvelik} A. O. Gogolin, A. A. Nersesyan and A. M. Tsvelik, {\it
Bosonization and Strongly Correlated Systems} (Cambridge Univ. Press,
Cambridge, 1998) and references therein.


\bibitem{ogata} M. Ogata and H. Shiba, Phys. Rev. B {\bf 41}, (1990) 2326.


\bibitem{haldane} F. D. M. Haldane, Phys. Rev. B {\bf 25}, (1982) 4925.



\bibitem{bra} S. Brazovskii, cond-mat/0401309 and references therein; 
D. Staresinic {\it et al.}, cond-mat/0509146. 


\bibitem{mon} S. Brazovskii, P. Monceau and F. Nad, Synthetic Materials 
{\bf 137}, (2003) 1331. 


\bibitem{bishop} H. Fehske {\it et al.} Physica B {\bf 359-361}, (2005) 699.


\bibitem{next} S. Lal and M. S. Laad, cond-mat/0511458; a comprehensively 
extended work is presented in cond-mat/0708.2159.



\bibitem{giamschulz} T. Giamarchi and H. J. Schulz, J. de Physique (Paris) 
{\bf 49}, (1988) 819. 


\bibitem{blum} G. Blumberg {\it et al.}, Science {\bf 297}, (2002) 584.

\bibitem{suzuki} M. Suzuki, Prog. Theor. Phys. {\bf 46}, (1971) 1337.

\bibitem{itzdrouffe}  C. Itzykson and J-M. Drouffe, {\it Statistical Field 
Theory} (Cambridge Univ. Press, Cambridge, UK, 1989), Vol.1.

\bibitem{irkhin} V. Y. Irkhin and A. A. Katanin, Phys. Rev. B {\bf 61}, 
(2000) 6757.

\bibitem{bocquet2} M. Bocquet, Phys. Rev. B {\bf 65}, (2002) 184415.

\bibitem{arrigoni} E. Arrigoni, Phys. Rev. B {\bf 61}, (2000) 7909; S. 
Biermann, A. Georges, A. Lichtenstein and T. Giamarchi, Phys. Rev. Lett. 
{\bf 87}, (2001) 276405.

\bibitem{delfino} G. Delfino, G. Mussardo and P. Simonetti, Nucl. Phys. B 
{\bf 473}, (1996) 469.

%
\bibitem{zamol} A. B. Zamolodchikov, Int. J. Mod. Phys. A {\bf 3}, 
(1988) 743.

\bibitem{vojta} M. Vojta, A. H\"ubsch and R. M. Noack, Phys. Rev. B 
{\bf 63}, (2001) 045105.


\bibitem{konik} R. M. Konik and Y. Adamov, cond-mat/0707.1160



\end{thebibliography}
\end{document}